\documentclass[12pt]{emulateapj}

\usepackage{yfonts}
\usepackage[T1]{fontenc}
\usepackage{ae,aecompl}

\usepackage{graphicx}	
\usepackage{amsmath}	
\usepackage{amssymb}	

\usepackage{microtype}
\usepackage{booktabs}
\usepackage{threeparttable}
\usepackage{tabularx}
\usepackage{xspace}

\usepackage{hyperref}

\newcommand{\project}[1]{\textsl{#1}}
\newcommand{\nustar}{\project{NuSTAR}\xspace}
\newcommand{\fermi}{\project{Fermi}\xspace}
\newcommand{\rxte}{\project{RXTE}\xspace}

\newcommand{\astrosat}{\project{Astrosat}\xspace}
\newcommand{\ixpe}{\project{IXPE}\xspace}

\begin{document}

\title[On The Statistical Properties of Cospectra]{On The Statistical Properties of Cospectra}

\author{D.~Huppenkothen\altaffilmark{1,2,3} and M.~Bachetti\altaffilmark{4}}

\altaffiltext{1}{Center for Data Science, New York University, 65 5h Avenue, 7th Floor, New York, NY 10003}
\altaffiltext{2}{Center for Cosmology and Particle Physics, Department of Physics, New York University, 4 Washington Place, New York, NY 10003, USA}
\altaffiltext{3}{DIRAC Institute, Department of Astronomy, University of Washington, 3910 15th Ave NE, Seattle, WA 98195}
\altaffiltext{4}{INAF-Osservatorio Astronomico di Cagliari, via della Scienza 5, I-09047 Selargius (CA), Italy}

\begin{abstract}
In recent years, the cross spectrum has received considerable attention as a means of characterising the variability of astronomical sources as a function of wavelength. While much has been written about the statistics of time and phase lags, the cospectrum has only recently been understood as means of mitigating instrumental effects dependent on temporal frequency in astronomical detectors, as well as a method of characterizing the coherent variability in two  wavelength ranges on different time scales. In this paper, we lay out the statistical foundations of the cospectrum, starting with the simplest case of detecting a periodic signal in the presence of white noise. This case is especially relevant for detecting faint X-ray pulsars in detectors heavily affected by instrumental effects, including \nustar, \astrosat\ and \ixpe. We show that the statistical distributions of both single and averaged cospectra differ considerably from those for standard periodograms. While a single cospectrum follows a Laplace distribution exactly, averaged cospectra are approximated by a Gaussian distribution only for more than $\sim\!\! 30$ averaged segments, dependent on the number of trials. We provide an instructive example of a quasi-periodic oscillation in \nustar\ and show that applying standard periodogram statistics leads to underestimated tail probabilities for period detection. We also demonstrate the application of these distributions to a \nustar\ observation of the X-ray pulsar Hercules X-1.

\end{abstract}








\section{Introduction}

Time series analysis is one of the primary ways to understand the physical properties of astronomical object in our universe, from exoplanets and stars to black holes and Active Galactic Nuclei (AGN). 
Fourier analysis, especially through the periodogram\footnote{We distinguish in this paper between the \textit{power spectrum}, which describes the process at the source generating variable time series, and the \textit{periodogram}, which denotes a realization of said power spectrum, i.e.\ the time series we actually observe. In line with the signal processing literature, we also use the term \textit{periodogram} or \textit{power spectral density} for the square of the absolute value of Fourier amplitudes derived from an observed light curve. We use the terms \textit{cospectrum} or \textit{cospectral density} to denote the real component of the cross-spectrum, i.e.\ the result of a multiplication of the Fourier amplitudes of one light curve with the complex conjugate of the Fourier amplitudes of a second light curve (see also Section \ref{sec:whitenoise_cospectra} for an exact definition)}, has long been used to find periodic and quasi-periodic signals as well as characterize the stationary stochastic processes often present in accreting systems. 
While in principle, the statistics of the periodogram is well understood and characterized in the literature \citep[e.g.][]{vanderklis1989}, the periodogram is often subject to instrumental effects like dead time that change its statistical properties and thus make statistical inference difficult in practice \citep[e.g.][]{Zhang+95}.

In the past years, the field of \textit{spectral timing} has enjoyed significant success by making it possible to combine both temporal and spectral information in a single model. Within this framework, the complex cross spectrum---defined as the Fourier transform of one time series with the complex conjugate of the Fourier transform of a second time series--holds a central position. The cross spectrum is commonly used to compute phase lags, which are easily converted to time lags and are central for understanding e.g.\ reverberation mapping in accreting black holes \citep[see][for a recent review]{uttley2014}. 

The real part of the cross spectrum, also named \textit{cospectrum}, has gained less attention, but can be just as useful. It has long been used to study gravity waves in the Earth's atmosphere \citep[e.g.][]{john2016}, models of the Martian atmosphere \citep[e.g.][]{wang2016}, the solar heliosphere \citep[e.g.][]{vigeesh2017}, surface elevation of arctic sea ice \citep[e.g.][]{ardhuin2016} and drifting snow \citep[e.g.][]{paterna2016} as well as surface gravity waves in beaches \citep[e.g.][]{fiedler2015} and eddy heat flux in the Earth's troposphere \citep[e.g.][]{wang2015,zurita-gotor2017}.

Within astronomy, and in particular the field of X-ray timing, it has recently been recognized as a powerful solution to two different problems. One problem is the reliable estimation of the variability contained in certain features of the power spectrum like quasi-periodic oscillations (QPOs) via the fractional rms amplitude. The rms amplitude crucially depends on reliable estimation of the Poisson noise level, which is sometimes difficult to calculate in practice. For instruments with two independent, identical detectors, the cospectrum of the light curves measured in each detector, respectively, will not contain any white noise contribution, since Poisson counting statistics are effects local to the instrument, and thus the observed data sets are independently sampled in each detector. This makes estimation of source-intrinsic variability much more reliable than comparable measurements using the periodogram.

The cospectrum is also effectively used as as one approach to mitigate instrumental effects, in particular an effect called \textit{dead time}. In many X-ray detectors, after detection of a photon there is a time interval during which the detector (or in imaging detectors the individual pixel) is blocked from detecting a second photon. This time interval is generally called \textit{dead time} (the detector is effectively ``dead'', and incoming photons will not produce a signal) and its characteristic time scale is set by the details of the detector and can range from very short ($2\mu\mathrm{s}$ in the \fermi\ Gamma-Ray Burst Monitor) to very long ($2.5\mathrm{ms}$ in \nustar). It leads to frequency-dependent changes in the mean and variance of the statistical distributions governing the periodogram, which cannot be mitigated by averaging periodograms of multiple segments.

Where in standard analysis, light curves of multiple detectors are summed before Fourier transforming the summed light curve, it is possible to instead Fourier-transform the signal of two independent detectors within the same instrument observing the same source strictly simultaneously in the same energy band. 
The resulting co-spectral densities will be less affected by dead time (see details in \citealt{Bachetti+15}), because the latter is introduced in each detector independently and tends to cancel out in the cospectrum.
This approach relevant for current X-ray missions carrying at least two identical detectors such as \nustar\ and \astrosat, but will also be relevant to future missions with multiple detectors like \ixpe. It has recently been used in \nustar\ studies of millisecond pulsars \citep{ferrigno2017}, Ultraluminous X-ray Sources \citep{bachetti2016} and X-ray binaries \citep{barriere2015,zoghbi2016,ingram2016,huppenkothen2017,stiele2017}. It is particularly important for instruments like \nustar\ which have a fairly long dead time and thus exhibit dead time effects in the periodogram at frequencies comparable to those where signals are expected, and at fairly low count rates.  
Similarly, the cospectrum of two time series taken with the same instrument, but in different wavelength ranges, can be used to characterize the coherent, phase-aligned variability in both time series as a function of frequency.

While much has been written on the subject of the statistics of cross spectra and time lags \citep[e.g.][]{epitropakis2016}, the derivation of cospectral statistics is notably absent from the astronomy literature, and most publications utilizing the approach of forming co-spectra from identical, but independent detectors observing the same source assume that either the $\chi^2_2$ distribution used for standard periodograms or a Gaussian distribution for averaged spectra is appropriate. That these assumptions are valid has not been shown until now. 

In this paper, we lay out the basic statistical distributions for detecting periodic and narrow quasi-periodic signals in the presence of detector noise (e.g.\ photon counting statistics) for both single and averaged cospectra. We show below that unlike for the periodogram, the statistical distribution for a single cospectrum reduces to a Laplace distribution, while the distribution for averaged cospectra is considerably more complex. We also find that for averaged cospectra consisting of more than $\sim$30 averaged individual segments, the assumption of a Gaussian distribution is indeed appropriate for single-trial tail probabilities and reduces computation overhead. However, for averaged cospectra of fewer segments, more stringent significance thresholds or large numbers of trials, the statistical distribution--and hence the derived $p$-values used for period detection--deviate significantly from a Gaussian distribution. We note that the results below hold only for cases where the signal to be detected is phase-aligned in both light curves. This must be true by definition for analyses utilizing data from independent, identical detectors observing the same source simultaneously (unless detectors are so far apart that light travel time effects might be an issue and photon arrival times are not barycentre-corrected to the gravitational centre of the solar system), but might not strictly be true for observations of the same source e.g.\ in different energy bands. We caution the reader that for signals that are not phase-aligned, the resulting power may decrease significantly and result in a non-detectable signal.

The paper is laid out as follows. In Section \ref{sec:single_cospectrum}, we derive the PDF of a single cospectrum, and show associated simulations and detection thresholds for period detection in Section \ref{sec:detectionthresholds}. Section \ref{sec:averaged_cospectra} extends the derivation to the common case where the cospectra of several time series are averaged. We show that the statistical distribution indeed changes from a Laplace distribution once multiple cospectra are averaged and becomes consecutively more Gaussian as a larger number of cospectra are included in the average in Section \ref{sec:averaged_detthres}. Finally, Section \ref{sec:nustarqpo} presents two real-world examples: the first uses simulated \nustar\ data of a quasi-periodic oscillation (QPO) as commonly found in accreting neutron star X-ray binaries. The second example comprises a real \nustar\ observation of the bright X-ray pulsar Hercules X-1. We end in Section \ref{sec:discussion} with a short discussion and conclusion. All figures and results are reproducible, and the associated code can be found online\footnote{\url{https://github.com/dhuppenkothen/cospectra-paper}}.
In a second, forthcoming paper, we will treat the considerably more complex case of cospectra where the time series consist of stochastic variability and show how to model the cospectrum in both a Maximum Likelihood and Bayesian framework.

For the reader looking for the statistical distributions of relevance, who may be only casually interested in the mathematical background, we point to Equations \ref{eqn:csdist} and \ref{eqn:averaged_pdf} for the probability density functions (PDFs) for a single cospectrum and averaged cospectrum, respectively, and Equations \ref{eqn:cospectrum_cdf} and \ref{eqn:averaged_cdf} for the cumulative distribution functions (CDFs) in both cases.

\section{The Statistical Distributions of Cospectral Densities}
\label{sec:whitenoise_cospectra}

In the following, we will consider the example of detecting a strictly periodic or very narrow quasi-periodic signal in the presence of simple white noise, as is commonly the case for example in pulsar searches in X-ray data. We assume that both the white noise and the periodic signal are wide-sense stationary (i.e. the mean and the autocovariance of the time series do not change with time) and the light curve is evenly sampled simultaneously in two identical detectors, or that photon arrival times are binned in intervals of equal length. While the white noise is also assumed to be strictly stochastic, the signal to be detected may be either deterministic (in the case of a strictly periodic signal) or stochastic (for a quasi-periodic oscillation). We then aim to reject a null hypothesis where the observed cospectral density at a given frequency can be explained by white noise alone.

\subsection{Statistical Distribution for a Single Cospectrum}
\label{sec:single_cospectrum}

Consider two independently distributed, evenly-sampled, constant, stationary time series,

\begin{eqnarray}
\mathbf{x} &=& \{x_k\}_{k=1}^N \nonumber \\
\mathbf{y} &=& \{y_k\}_{k=1}^N \nonumber
\end{eqnarray}

\noindent with $N$ data points taken at simultaneous time intervals $\{t_k\}_{k=1}^N$ with a constant time resolution $\Delta t$ and a total duration $T = N\Delta t$. Assume for simplicity that the measurements $x_k$ and $y_k$ are normally distributed, such that 

\begin{eqnarray}
x_k &\sim& \mathcal{N}(\overline{x}, w_x^2)  \nonumber \\
y_k &\sim& \mathcal{N}(\overline{y}, w_y^2) \nonumber
\end{eqnarray} 

\noindent with means $\overline{x}, \overline{y}$ and variances $w_x^2, w_y^2$. The data points in the time series $\mathbf{x}$ and $\mathbf{y}$ can be expressed in terms of a Fourier series,

\begin{eqnarray}
x_k & = & \frac{1}{N} \sum_{j}{\mathcal{F}_x(j)} \nonumber \\
x_y & = & \frac{1}{N} \sum_{j}{\mathcal{F}_y(j)}
\end{eqnarray}

\noindent where

\begin{eqnarray}
\mathcal{F}_x(j) &= & \frac{1}{2} (A_{xj} - i B_{xj}) e^{-i\left( \frac{2 \pi j t}{T} \right)} \\
\mathcal{F}_y(j) &= & \frac{1}{2} (A_{yj} - i B_{yj}) e^{-i\left( \frac{2 \pi j t}{T} \right)} \, .
\end{eqnarray}

\noindent Here, $i = \sqrt{-1}$, and $A_{xj}, A_{yj}$ and $B_{xj}, B_{yj}$ describe the real and imaginary parts of the Fourier amplitudes, respectively (for a pedagogical introduction into Fourier analysis, see \citealt{vanderklis1989}). We restrict $\mathcal{F}_x(j)$ and $\mathcal{F}_y(j)$ to frequencies between $\nu_{j=0} = 1/T$ and the Nyquist frequency $\nu_{j=N/2} = 1/(2\Delta t)$.
The complex cross spectrum is then calculated by multiplying the Fourier transform of light curve $\mathbf{x}$ with the complex conjugate of the Fourier transform of light curve $\mathbf{y}$ (\citealt{vaughan1997,nowak1999}, see also \citealt{uttley2014} for a recent review of spectral timing techniques):

\begin{eqnarray}
\mathcal{F}_x(j) \mathcal{F}_y^*(j) & = & \frac{1}{2} (A_{xj} - i B_{xj}) e^{i \frac{2 \pi j t}{T}} \frac{1}{2} (A_{yj} + i B_{yj}) e^{i \frac{-2 \pi j t}{T}}\nonumber \\ 
		     & = & \frac{1}{4} [ (A_{xj}A_{yj} + B_{xj}B_{yj}) + \\\nonumber
		     & &  i (A_{xj}B_{yj} - A_{yj}B_{xj}) ]
\end{eqnarray}

Note that for strictly real-valued time series, as light curves in astronomy always are, $A_j = A_{-j}$ and $B_j = - B_{-j}$, such that 

\begin{equation}
\mathcal{F}_x(j) \mathcal{F}_y^*(j) = \frac{1}{2} \left\{ (A_{xj}A_{yj} + B_{xj}B_{yj}) + i (A_{xj}B_{yj} - A_{yj}B_{xj}) \right\} \, .
\end{equation}

\noindent The real part of this equation is the cross-spectral equivalent of the power spectral density, also called the cospectrum:

\begin{equation}
C_j = \frac{1}{2}  (A_{xj}A_{yj} + B_{xj}B_{yj}) \, .
\end{equation}

\noindent For evenly sampled, normally distributed light curves (and indeed for uncertainties coming from a wide range of statistical distributions, including the Poisson distribution, if $N$ is large), the real and imaginary amplitude components are distributed as $A_{xj}, B_{xj} \sim \mathcal{N}(0, \sigma_x^2)$ with $\sigma_x =  \sqrt{\frac{\sum_{k=1}^{N}{x_k}}{2}}$ and  $A_{yj}, B_{yj} \sim \mathcal{N}(0, \sigma_y^2)$ with $\sigma_y = \sqrt{\frac{\sum_{k=1}^{N}{y_k}}{2}}$. Note that these distributions for the Fourier amplitudes only hold if the underlying process producing the observations is stochastic and wide-sense stationary. This includes many processes commonly observed in astrophysical sources such as white noise observed from a constant background, as well as red noise processes such as shot noise and other (broken) power-law power spectra often seen in AGN, and quasi-periodic oscillations with stochastic variations in amplitude and period of the observed signal, commonly observed in black hole X-ray binaries. For strictly deterministic processes (e.g.\ strictly periodic variations), the distributions of the Fourier amplitudes will not be centered on $\mu = 0$, and the distributions below will not be correct.

For standard periodograms, $A_{xj} = A_{yj}$ and $B_{xj} = B_{yj}$, and the power spectral density reduces to $P_j = \frac{1}{2} (A_j^2 + B_j^2)$, which is well-known to follow a $\chi^2$ distribution with $2$ degrees of freedom. 

Because this condition is not fulfilled for cospectra, we need to derive the probability distribution of the sum of the products over \textit{independent} Gaussian random variables. The probability distribution of the product of two random variables\footnote{We continue the following derivation using the amplitudes $A_j$, but the same arguments apply exactly to the imaginary amplitudes $B_j$.} $Z = A_{xj}A_{yj}$ is called the product distribution, defined as 

\begin{eqnarray}
p_Z(z) & = & \int_{-\infty}^{+\infty}{p_X(x) p_Y(\frac{z}{x}) \frac{1}{|x|} dx} \nonumber \\
	   & = &  \int_{-\infty}^{+\infty}{\frac{1}{2 \pi \sigma_x \sigma_y} \exp{-\frac{x^2}{2\sigma_x^2}} \exp{-\frac{(z/x)^2}{2\sigma_y^2}} \frac{1}{|x|} dx} \label{eqn:pz} \, .
\end{eqnarray}

\noindent It can be shown \citep{watson1922,wishart1932} that the integral in Equation \ref{eqn:pz} above can be reduced to

\begin{equation}
\label{eqn:bessel}
P_Z(z) = \frac{K_0\left( \frac{|z|}{\sigma_x \sigma_y}\right)}{\pi \sigma_x \sigma_y} \, ,
\end{equation}

\begin{figure*}
\begin{center}
\includegraphics[width=\textwidth]{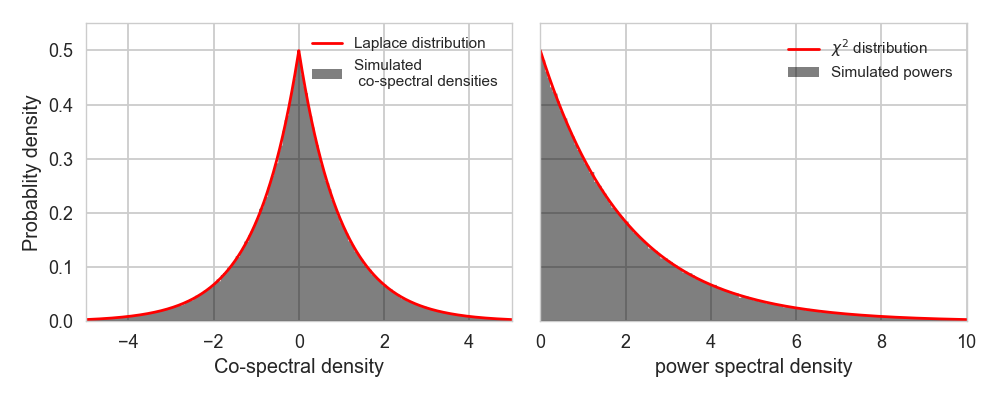}
\caption{Distribution of Leahy-normalized cospectral densities (left) and power spectral densities (right), respectively, for the simulated data. In dark grey, we show fine-grained histograms of the simulated densities. In red we plot the theoretical probability distribution the simulated densities should follow: A Laplace distribution with $\mu=0$ and $\sigma=2$ for the cospectral densities and a $\chi^2$-distribution with $2$ degrees of freedom for the power spectral densities. The simulated densities adhere very closely to the theoretical predictions.}
\label{fig:csdist}
\end{center}
\end{figure*}

\noindent where $K_0(x) = \int_{0}^{+\infty}{\frac{\cos{(xt)}}{\sqrt{t^2 + 1}} dt}$ is the Bessel function of the second kind of order $0$. We can now use this result to derive the probability density function of $C_j$. In particular, we find that both random variables $Z_j = A_{xj} A_{yj}$ and $Q_j = B_{xj} B_{yj}$ follow the Bessel distribution defined in Equation \ref{eqn:bessel}. Our task is therefore to find the PDF of the sum of two Bessel distributions. The PDF of this sum requires the convolution of the PDFs of each individual random variable being summed. This convolution is difficult to calculate directly for the Bessel distribution defined in Equation \ref{eqn:csdist} above. We instead consider the \textit{moment-generating function} of the PDF, generally defined as 

\begin{equation}
M_Z(t) := \mathbb{E}[e^{tZ}] \, 
\end{equation}

\noindent for a random variable $Z$. Consider the sum of any two independent random variables, $S = Z + Q$. While the PDF of $S$ can be found via the convolution of the individual PDFs, it is often simpler to consider the moment-generating function, where the convolution reduces to a simple multiplication operation:

\begin{equation}
M_S(t) = M_Z(t) M_Q(t) \, .
\end{equation} 

\noindent The moment-generating function of the Bessel distribution in Equation \ref{eqn:bessel} above is, in the general case \citep{seijas2012} where the means $\mu_x$ and $\mu_y$ are non-zero and the random variables may have unequal variances $\sigma_x \neq \sigma_y$: 

\begin{equation}
M_Z(t) = \frac{\exp{\left( \frac{t\mu_x \mu_y + 0.5 (\mu_y^2 \sigma_x^2 + \mu_x^2 \sigma_y^2) t}{1 - t^2 \sigma_x^2 \sigma_y^2} \right)}}{\sqrt{1 - t^2 \sigma_x^2 \sigma_y^2}}\, ,
\end{equation}
but since $\mu_x = \mu_y = 0$ for the Fourier amplitudes of a stationary stochastic process, this reduces to
\begin{equation}
M_Z(t) =  \frac{1}{\sqrt{1 - t^2 \sigma_x^2 \sigma_y^2}}  \label{eqn:mgf},
\end{equation}

Thus, the moment-generating function for the sum of $Z_j$ and $Q_j$ becomes

\begin{equation}
M_C(t) = M_Z(t) M_Q(t) = \frac{1}{1 - t^2 \sigma_x^2 \sigma_y^2} \, .
\label{eqn:mgt_c}
\end{equation}

\noindent We note that the Laplace distribution is defined as 

\[
p_{\mathrm{Laplace}}(x | \mu, b) = \frac{1}{2b} \exp{\left(-\frac{|x - \mu|}{b} \right)}
\]

\noindent and its moment-generating function as

\[
M_\mathrm{Laplace}(t) = \frac{e^{t\mu}}{1 - b^2 t^2} \, .
\]

Comparing this last equation with Equation \ref{eqn:mgt_c}, we find that that Equation \ref{eqn:mgt_c} is equal to the moment-generating function of the Laplace distribution with $\mu = 0$ and $b = \sigma_x \sigma_y$, and hence the cospectral densities follow a Laplace distribution:

\begin{equation}
p(C_j | 0, \sigma_x\sigma_y) = \frac{1}{\sigma_x \sigma_y} \exp{\left(- \frac{|C_j|}{\sigma_x\sigma_y} \right)} 
\label{eqn:laplace}
\end{equation}

\noindent with

\begin{equation}
\sigma_x =  \sqrt{\frac{\sum_{k=1}^{N}{x_k}}{2}} \;\;\; \mathrm{and} \;\;\; \sigma_y =  \sqrt{\frac{\sum_{k=1}^{N}{y_k}}{2}} \, .
\label{eqn:csdist}
\end{equation}

\subsubsection{Detection Thresholds}
\label{sec:detectionthresholds}

Detection thresholds for cospectra will generally be different from those of classical periodograms, because the Laplace distribution tends to be narrower than the equivalent $\chi^2_2$ distribution for single periodograms. To show how the distributions and the corresponding detection thresholds differ, we simulated simple Poisson-distributed light curves. First, we simulated two light curves with a duration of $10\,\mathrm{s}$ and $10^6$ data points each, corresponding to a time resolution of $10^{-5}\,\mathrm{s}$. The light curves have an identical mean count rate of $10^{6} \, \mathrm{counts/s}$, corresponding to $10$ counts per bin. In order to simulate typical measurement uncertainties in X-ray detectors, we sampled from a Poisson distribution for each time bin with a rate parameter $\lambda = 10$, corresponding to the average counts per bin.

\begin{figure}
\begin{center}
\includegraphics[width=9cm]{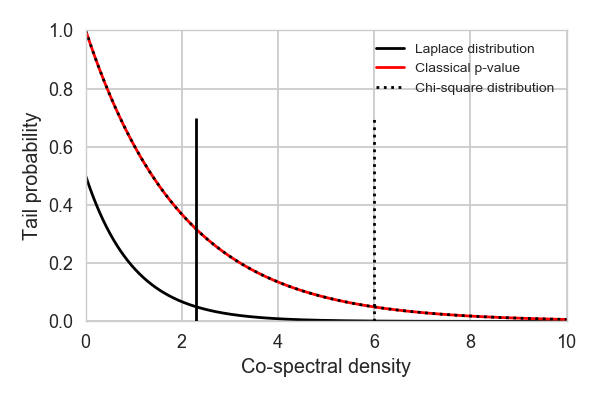}
\caption{Tail probabilities for the Laplace and $\chi^2$ distributions, respectively. The tail probability, or survival function, is defined as $SF(x) = 1 - CDF(x)$. The tail probability measures the probability of observing a value $x\geq X$, and is often used for detecting periods in periodograms. For the power spectral densities, we plot both the theoretical prediction for the survival function based on the $\chi^2$ distribution (black dashed line), as well as the corrected distribution for periodograms derived in \citet{groth1975} (red solid line). For illustrative purposes, we show a single-trial $95\%$ detection threshold for the Laplace distribution (black solid vertical line) and the $\chi^2$ distribution (black dashed vertical line).}
\label{fig:survival}
\end{center}
\end{figure}

We then calculated both the cospectrum of the two light curves and the periodogram of only the first light curve for comparison. For simplicity, both spectra were computed in Leahy normalization \citep{leahy1983}, which is typically used when searching for (quasi-)periodic signals in time series.
In order to normalize the cospectrum correctly, we used $2/\sqrt{N_{\mathrm{ph}, x}N_{\mathrm{ph}, y}}$, where $N_{\mathrm{ph}, x}$ and $N_{\mathrm{ph}, y}$ are the number of photons of light curves $\mathbf{x}$ and $\mathbf{y}$, respectively, as prescribed by \citet{Bachetti+15}.
In this normalization the densities are distributed as $\chi^2_2$ exactly for the periodogram, and following a Laplace distribution with $\mu=0$ and $\sigma = 1$ for the cospectrum. In Figure \ref{fig:csdist}, we plot the resulting distribution of densities. While the periodogram is only defined for positive values, the Laplace distribution is symmetric around zero, and in general the cospectrum will comprise both positive and negative densities. It is also immediately visible from Figure \ref{fig:csdist} that the probability of obtaining a certain (positive) noise power is always lower for the Laplace distribution than for the $\chi^2$ distribution. In practice, this implies that using the latter where the former is appropriate, we may miss significant periodic signals, because we assume them to be weaker than they are in reality. To demonstrate this, we plot the survival function in Figure \ref{fig:survival}. The survival function, defined in terms of the CDF as $SF(x) = 1 - CDF(x)$, encodes the tail probability of seeing at least a value $x \geq X$. This tail probability is often considered to be the $p$-value of rejecting the null hypothesis that a certain candidate for a periodic signal could be reasonably produced by the noise powers. The CDF for the Laplace distribution with $\mu=0$ is defined as 

\begin{equation}
\label{eqn:cospectrum_cdf}
F_{C_j}(x)) = 
  \begin{cases} 
   \frac{1}{2} \exp{\left(\frac{C_j}{\sigma_x \sigma_y}\right)}  & \text{if } C_j < 0 \\
    1 - \frac{1}{2} \exp{-\left( \frac{C_j}{\sigma_x \sigma_y} \right)}   & \text{if } C_j \geq 0
  \end{cases}
\end{equation}

\begin{figure}
\begin{center}
\includegraphics[width=9cm]{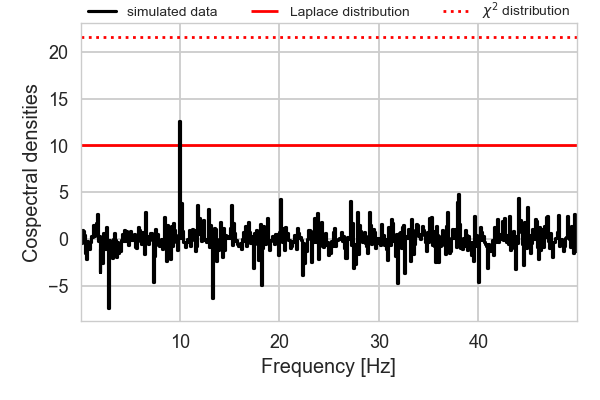}
\caption{Cospectrum of two simulated light curves, each with a constant continuum flux of $10$ counts per bin and a periodic signal at $10\,\mathrm{Hz}$. The latter is clearly visible in the cospectrum. We also show the 99\% detection threshold, corrected for a number of trials equal to the number of spectral bins, assuming Laplace-distributed data (red solid line) and $\chi^2$-distributed data (red dashed line). When the latter distribution is assumed, the periodic signal would not be considered a significant detection, because the $\chi^2$ distribution produces a wider distribution of densities. Applying the correct Laplace distribution, however, allows for the detection of weaker signals.}
\label{fig:cs_sim}
\end{center}
\end{figure}

\noindent Much like the PDF, the tail probability is always smaller for the Laplace distribution, indicating that for a given candidate signal, the $p$-value for rejecting the null hypothesis will be stronger than for $\chi^2$-distributed variables. To reinforce this statement, we again simulated two light curves, each again with a duration of $10\,\mathrm{s}$, but this time with only $1000$ data points for simplicity and speed, and a time resolution of $0.01\,\mathrm{s}$. For this simulation, we assumed a mean count rate of $1000\,\mathrm{counts/s}$ or $10$ counts per bin, and additionally introduced a sinusoidal signal with a period of $0.1\,\mathrm{s}$ and a fractional rms amplitude of $a_\mathrm{frac} = 0.055$. Again, this template was used to produce two Poisson-distributed light curves with a rate parameter equal to the number of counts in each bin as defined by the flat continuum and the periodic signal. In Figure \ref{fig:cs_sim}, we show the cospectral densities along with trial-corrected $0.99$ detection thresholds for both the Laplace and $\chi^2$ distribution. If the densities are assumed to follow a $\chi^2$ distribution, as for the periodogram, the candidate at $10 \,\mathrm{Hz}$ would be discounted at the $99\%$ detection threshold, whereas correctly applying the Laplace distribution yields a correct rejection of the null hypothesis at the same detection threshold.

Note that for light curves affected by dead time, the resulting cospectrum will still follow the Laplace distribution above, but with a variable variance that changes as a function of frequency \citep{Bachetti+15}. In practice, the cospectrum can be straightforwardly corrected for this effect using the differences in Fourier amplitudes derived from the light curves of two detectors (the \textit{Fourier Amplitude Difference} (FAD) technique), and \citet{bachetti2017} show that the corrected cospectrum will closely follow the Laplace distribution derived here, allowing for unbiased significance tests for periodicity detection.

\begin{figure*}
\begin{center}
\includegraphics[width=\textwidth]{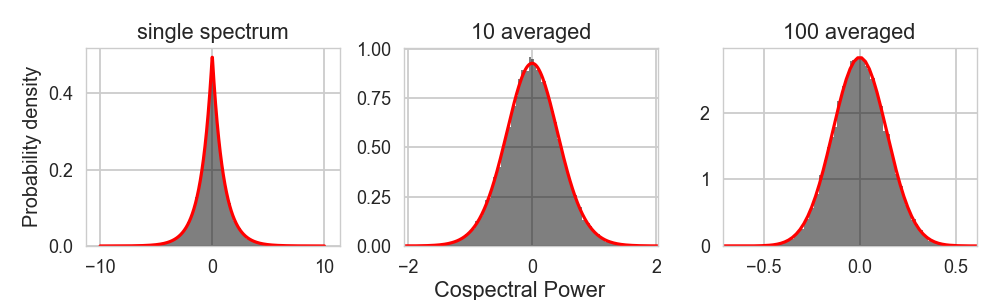}
\caption{Histogram of cospectral densities (grey) and the theoretical expectation (red) for three different cases. In the right panel, we show the distribution of densities for a single cospectrum, with its expected Laplace distribution from Equation \ref{eqn:laplace}, in the middle and right panel cospectral densities for averaging 10 (middle) and 100 (right) individual cospectra together. In the middle panel, the theoretical expectation of the sampling distribution is given by Equation \ref{eqn:averaged_pdf}, in the right panel by a Gaussian distribution with $\sigma=\sqrt{2/101}$.}
\label{fig:avg_dist}
\end{center}
\end{figure*}

\subsection{Averaged Cospectra}
\label{sec:averaged_cospectra}

The $\chi^2$ distribution used for periodograms has the simple property that sums of $\chi^2$-distributed variables again follow the same distribution, with a different number of degrees of freedom. The same is not true for the Laplace distribution. For $n$ independent and identically distributed (i.i.d.) random variables distributed following a standard Laplace distribution with a mean of $\mu = 0$ and a width of $b = 1$, the distribution of the sums of these random variables can be derived using the fact that a single Laplace random variable $X$ can be rewritten as the difference of two exponential random variables, 

\[
X = Z - Z' \; ,
\]
\noindent and thus for $n$ summed random Laplace random variables,

\begin{equation}
T = \sum_{i=1}^{n} X_i = \sum_{i=1}^{n}Z_i - \sum_{i=1}^{n} Z'_i = G_1 - G_2 \, ,
\end{equation}

\noindent where $G1$ and $G2$ are i.i.d.\ standard gamma random variables with a distribution $g(x) = \frac{1}{\Gamma(\nu)} x^{\nu-1} e^{-x}$ and a shape parameter $\nu = n$. For the full derivation of the density, we refer the reader to \citet{kotz2001} and simply state the end result for the PDF for $n$ averaged standard Laplace random variables, $\overline{X}_n$ (see \citealt{kotz2001}, Equations 2.3.25 and 2.3.26):

\begin{equation}
f_{\overline{X}_n}(x) = \frac{n e^{-|nx|}}{(n-1)! 2^n} \sum_{j=0}^{n-1} \frac{(n-1+j)!}{(n-1-j)! j!} \frac{|nx|^{n-1-j}}{2^j} \; \;\;, x \in R \; .
\label{eqn:averaged_pdf}
\end{equation}

For practical purposes, evaluating this PDF for averaged spectra above $n\sim 85$ is difficult numerically, because the factorials and exponents in the sum become very large and small, respectively. However, as we will show in Section \ref{sec:averaged_detthres} below, we expect that for large $n$, the Central Limit Theorem implies that the PDF of averaged cospectral densities tends towards a normal distribution. We find that in practice, when $n \gtrsim 30$, detection thresholds derived from Equation \ref{eqn:averaged_pdf} provide only a negligible difference over that derived from a normal distribution $N(0, \sqrt{2/(n+1)}$, depending on the significance threshold required and the number of trials. 

In order to derive tail probabilities useful for hypothesis testing, we require the CDF rather than the PDF. In order to correctly account for the absolute values in the PDF, we split the CDF into two parts: a case where $x < 0$ and a case where $x \geq 0$. The integral $F_{\overline{X}_n}(x) = P ( X \leq x) = \int_{\infty}^{x} f_{\overline{X}_n}(t) dt$ then becomes:

\begin{equation}
\label{eqn:averaged_cdf}
F_{\overline{X}_n}(x)) = 
  \begin{cases} 
   \sum_{j=0}^{n-1} D \frac{1}{n}(2\Gamma(-j+n) - \gamma(-j+n, nx)) \;, \;x \geq 0 \\
     \sum_{j=0}^{n-1} D  \frac{1}{n}\gamma(-j+n, -nx) \;,\; x < 0
  \end{cases}
\end{equation}

\noindent where $\Gamma(l)= (l-1)!$ is the gamma function, $\gamma(l+1, m) = l\gamma(l,m) - l^m e^{-m}$ is the incomplete upper gamma function, and the pre-factor constant $D$ is defined as 

\[
D = \frac{n(n-1+j)!}{(n-1-j)! j! (n-1)! 2^{n+j}} \, .
\]

\noindent As laid out in Section \ref{sec:detectionthresholds}, the tail probability can easily be calculated via the survival function, $\mathrm{SF}(x) = 1 - \mathrm{CDF}(x)$.

\begin{figure*}
\begin{center}
\includegraphics[width=\textwidth]{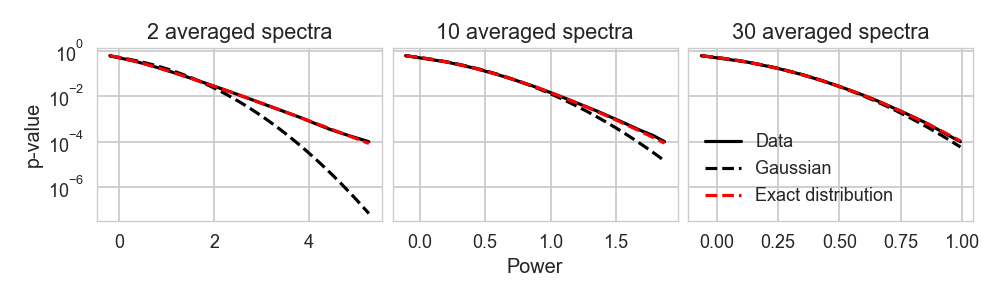}
\caption{Predictions for single-trial $p$-values as a function of cospectral density, for the simulated data sets (black solid line) compared to the theoretically predicted $p$-values using the CDF derived in Equation \ref{eqn:averaged_cdf} (red dashed line) as well as the survival function of a Gaussian distribution (black dashed line). The exact distribution matches the empirically derived tail probabilities from simulations within the uncertainties of the simulations. For the case of two averaged cospectra (left panel), a Gaussian approximation is obviously the wrong choice, and will lead to vastly overestimated significances, because the Gaussian PDF drops off much more sharply than the exact distribution. However, as more cospectra are averaged, the resulting distribution becomes more and more similar to that of a Gaussian (middle panel), and for $\sim$30 averaged cospectra, a Gaussian survival function yields a reasonably good approximation to the tail probabilities one would derive from simulations (right panel), up to $p \approx 10^{-4}$.}
\label{fig:avg_pvalue}
\end{center}
\end{figure*}

\subsubsection{Detection Thresholds}
\label{sec:averaged_detthres}

In order to show the way the probability distribution changes as a function of averaged cospectra, we simulate light curves of $10^{5}$ data points and a mean count rate of $100\,\mathrm{counts}/\mathrm{s}$ consisting of pure white noise. We compute $n$ such light curves and average their cospectral densities in order to show the distribution of those densities compared to the expected probability distributions. In Figure \ref{fig:avg_dist}, we show the simulated distribution of Leahy-normalized power spectral densities, along with the distributions that describe them. For a single, non-averaged spectrum, we use the Laplace PDF described in Equation \ref{eqn:laplace}. When averaging $10$ cospectra, we use Equation \ref{eqn:averaged_pdf} and show that the theoretical predictions agree with the simulated densities . Finally, for an averaged cospectra consisting of 100 individual light curves, Equation \ref{eqn:averaged_pdf} becomes difficult to compute numerically, and we use a Gaussian PDF instead, which describes the distribution of simulated densities well. 

In order to assess the effect on the $p$-values derived from averaged cospectra, we calculate the tail probabilities for the simulated data sets and compare them with the theoretically expected survival function as defined in Equation \ref{eqn:averaged_cdf} as well as a simple Gaussian distribution (Figure \ref{fig:avg_pvalue}). Similar to the results derived by \citet{balakrishnan1986}, we find that for cospectra of more than 30 averaged light curves, a Gaussian distribution yields a reasonably good approximation to the true distribution up to $p \approx 10^{-4}$ with lower overhead. Note, however, that this holds for single-trial probabilities. In general, one will wish to correct for calculating the significance of multiple trial frequencies, requiring the use of more stringent significance threshold. As shown in Figure \ref{fig:avg_pvalue}, the tail probabilities diverge as a function of power, and thus the Gaussian approximation will increasingly overestimate the significance of the signal the higher the threshold is set. Depending on the number of trials used, it is hence advisable to use Equation \ref{eqn:averaged_cdf} as long as it remains numerically stable (depending on implementation, up to $\sim$85 averaged cospectra).

\begin{figure*}
\begin{center}
\includegraphics[width=\textwidth]{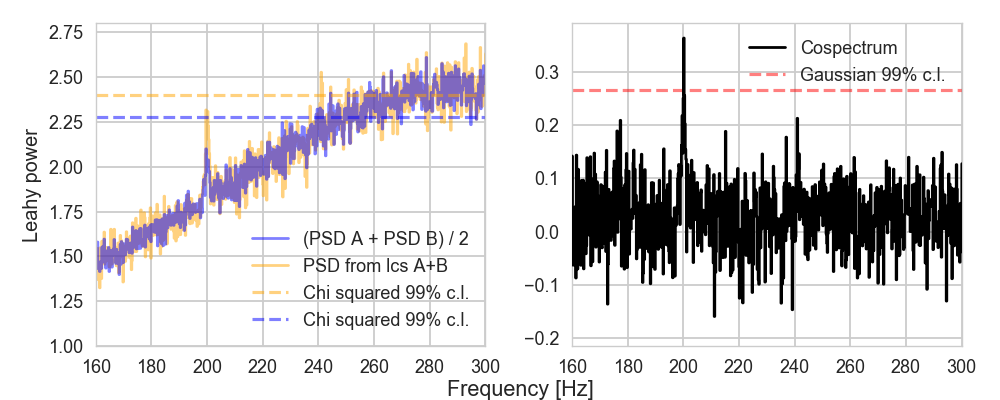}
\caption{Comparison of the averaged periodograms from the two detectors (blue), the periodogram obtained by the sum of the light curves (yellow; both left panel) and the cospectrum (right panel), for $3000\,\mathrm{s}$ of synthetic \nustar data with an incident count rate of $200 \mathrm{counts}\mathrm{s}^{-1}$ and a strong QPO at $200\,\mathrm{Hz}$. 
The QPO has an rms of $\sim$9\% and a high Q factor of $\sim$40. The shape of the periodogram is strongly distorted from the expected flat power spectrum centred on a constant value of $2$, with deviations of more that $0.5$ in the mean level of the densities common. In dashed lines, we also show the expected (trial-corrected) $99\%$ confidence level for the periodograms (in the same colour, left panel) as well as the cospectrum (red, right panel). The signal is detected in both the periodogram of the combined light curves as well as the cospectrum. However, because the white noise level is variable in the periodogram and the power spectrum dips below the expected level of $2$ around the frequency of the QPO, the signal is not detected as significantly as it should have been if no dead time was present. Assessing the significance is much more straightforward in the cospectrum, which retains a flat baseline.
}
\label{fig:qpo}
\end{center}
\end{figure*}
\section{Caveats}

All caveats applying more generally to Fourier analysis and to periodograms specifically also apply to cospectra. In particular, as for the $\chi^2$ distribution used for period searches in periodograms, the distributions derived here assume that the underlying process is wide-sense stationary and the light curves are simultaneously and evenly sampled. These conditions are typically met in most observations taken with X-ray timing instruments like \nustar\ and \rxte\ for problems such as timing of X-ray binaries, AGN and X-ray pulsar searches, but this may not be the case for e.g.\ optical observations of main-sequence or binary stars. Note that we also implicitly assume that the time scales of interest are much shorter than outburst timescales in X-ray binaries or changes in the X-ray background in pulsar searchers, both of which may not be stationary.

Non-stationarity (as is e.g.\ observed in flares) will impose a window function onto the cospectrum and shift the mean of the Fourier amplitudes, thus invalidating the assumptions above (for an illustrative example of this effect and its implications on period searches in periodograms of magnetar bursts, see \citealt{huppenkothen2013}). Uneven sampling, on the other hand, introduces covariances between adjacent frequencies and renders many of the statistical assumptions underpinning period searches using both in the periodogram and the cospectrum invalid. For unevenly sampled data, other methods that do not depend on a regular sampling pattern must be employed (for example the Lomb-Scargle periodogram \citep{lomb1976, scargle1982} or the Bayesian period search in \citet{gregory1992}) and dead-time effects must be forward-simulated.

An additional assumption is the phase-alignment of the periodic signal to be detected in both light curves. This is always true when the same source is observed simultaneously in the same energy range in independent, but identical detectors on the same spacecraft, as is common practice for instruments like \nustar, \astrosat\ and \rxte. Because the phase of the periodic signal can more generally be energy-dependent, we caution the reader to be careful when producing cospectra of light curves taken for example in different energy ranges. Similarly, the phase of a quasi-periodic signal can shift with times, thus observations at different points in time, or of different sources must be analyzed similarly carefully. In all these cases, period searches will only be sensitive to signals that are phase-aligned, and will in the worst case be undetectable if the phase-shift is 90 or 270 degrees.

In realistic applications, in particularly where dead time is a major concern, the cospectrum will show a frequency-dependent variation in the local variance, whose strength depends positively on the overall count rate of the object observed. It is imperative that this should be corrected before using the distributions used here by applying the Fourier Amplitude Differencing (FAD) technique \citep{bachetti2017}. This method, while powerful, is not without caveats. In particularly, FAD-corrected cospectra tend to overestimate the integrated rms when both count rate and variability amplitude are both very high. This should be accounted for when deriving estimates of the fractional rms amplitude.

\section{Example: \nustar}
\label{sec:nustarqpo}

In the following, we will consider two more realistic examples in turn. First, we will consider a more realistic simulation of a QPO as expected to be observed in a typical \nustar\ observation. Subsequently, we present an example of real \nustar\ data containing a coherent pulsed signal from the X-ray neutron star source Hercules X-1.

\begin{figure*}
\begin{center}
\includegraphics[width=\textwidth]{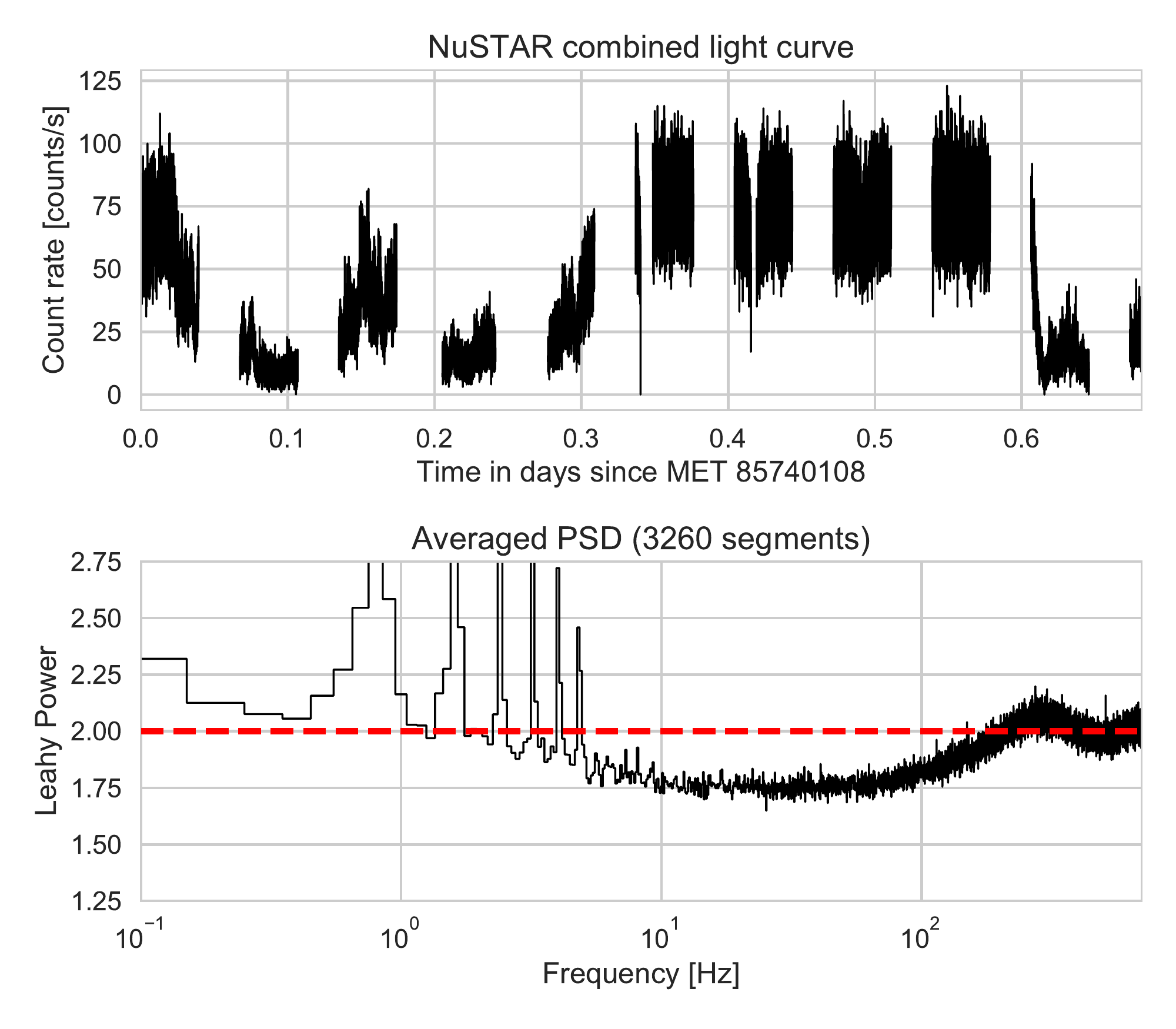}
\caption{Upper panel: Light curve of Her X-1 observed with \nustar, observation ID 30002006002. The observation totals $32.67\,\mathrm{ks}$, the light curve presented here is produced by combining event data from both FPMA and FPMB detectors. Gaps in the light curve are caused by the $96.8 \,\mathrm{m}$ orbit around Earth. Lower panel: averaged periodogram of $3260$ $10\,\mathrm{s}$-duration segments. The bright pulsar produces a very strong periodic signal at $\nu_\mathrm{rot} = 0.806\,\mathrm{Hz}$ ($P_\mathrm{rot} = 1.24\,\mathrm{s}$) and at the signal's harmonics. In order to show the effects of dead time on the underlying detector noise, we have zoomed into the noise region, such that the peaks produced by the pulsar extend beyond the upper edge of the plot. At the lowest frequencies, the periodogram deviates from the expected $\chi^2$ distribution centered on $\mu = 2$ because of longer-term, aperiodic variability in the source itself. At frequencies above a $\sim$5 Hz or so, dead time strongly modulates the noise powers into an oscillatory pattern, indicating that the use of standard statistical distributions commonly used to find periodic or quasi-periodic signals will lead to biased results.
}
\label{fig:herx1_lc}
\end{center}
\end{figure*}

\subsection{A Simulated QPO in \nustar}

In order to show the difference of the detection limits with the cospectrum and the power spectrum, we show how a QPO at 200 Hz from a very bright source would appear in \nustar. 
We simulate a light curve of $T=3000\mathrm{s}$ duration with a time resolution of $\delta t = 0.5 \,\mathrm{ms}$ and an average count rate of $200 \,\mathrm{counts}\,\mathrm{s}^{-1}$. To this constant background we add a quasi-periodic oscillation with a period of $5\,\mathrm{ms}$, a fractional rms amplitude of $f_\mathrm{rms} = 0.15$ and phases randomized using a normal distribution with a mean of $0$ and a width of $\sigma_\mathrm{qpo} = 0.01$. After producing this light curve, the procedure is similar to that followed by \citet{Bachetti+15}. We simulated photon events using rejection sampling from this light curve using the software package \textit{stingray}\footnote{\url{https://github.com/StingraySoftware/stingray}}, running the function \texttt{stingray.Eventlist.simulate\_times()} twice in order to produce two light curves that are statistically independent, but have the same signal and properties, as we would expect from an instrument with two independent detectors observing the same object. Subsequently we simulated variable deadtime for both light curves with an average time scale of $2.5 \,\mathrm{ms}$ as commonly seen in \nustar\ data \citep{Bachetti+15} using \textit{HENDRICS}\footnote{\url{https://github.com/StingraySoftware/HENDRICS}} \citep{bachetti2015b}. We then produced the periodogram of the summed light curves, the averaged periodogram of the two individual light curves, and the cospectrum of the two light curves. Note that in all three cases we produced averaged periodograms and co-spectra by splitting the light curves into 600 segments of $5\,\mathrm{s}$ length each in order to suppress the variance in the powers and show the effects of dead time more clearly. As mentioned above, cospectra with dead time are subject to frequency-dependent changes in variance, and we thus corrected the simulated cospectra using the FAD technique.

The results are shown in Figure~\ref{fig:qpo}. While in all three cases, the QPO is clearly visible, the two periodograms show strong deviations from the expected flat power spectrum. The shape is distorted and requires a precise model of the non-linearly increasing baseline with a non-linearly increasing rms. While in principle, the periodogram of the combined light curves would have a higher significance by a factor of $\sqrt{2}$, the modeling requirements complicate the calculation of the significance of the QPO. 
The baseline of the cospectrum, conversely, is not distorted by dead time, and requires only an estimate of the \textit{local} rms in order to calculate the significance of the QPO using Equation~\ref{eqn:averaged_cdf}: it is sufficient to \textit{multiply the cospectrum around the feature by an estimate of the local standard deviation of the white noise} (which is 1 in the ideal case) to use the equations above with no modifications.

\subsection{\nustar\ Observations of Her X-1}

Hercules X-1 (Her X-1) is a well-studied persistent X-ray binary pulsars \citep{tananbaum1972} in an X-ray binary comprising the neutron star itself and a stellar companion HZ Herculis \citep{davidsen1972,forman1972,bahcall1972} with a mass of $\sim$$2.2 \,\mathrm{M}_\odot$ \citep{reynolds1997,leahy2014} whose type varies between late-type A and early-type B with orbital phase \citep{anderson1994,cheng1995} . The neutron star spins with a period of $P = 1.23\,\mathrm{s}$ \citep{tananbaum1972} and the system overall exhibits an orbital period of $P_\mathrm{orb} = 1.7\,\mathrm{days}$ \citep{bahcall1972}, along with super-orbital variations on a $\sim$35-day timescale \citep{giacconi1973,scott1999,igna2011}.

Her X-1 has been observed by the Nuclear Spectroscopic Telescope Array (\nustar) multiple times. 
For this work, we considered the observation taken from UT 2012-09-19 to UT 2012-09-20, one of those used by \citet{fuerst13} to characterize the cyclotron resonance scattering features in the spectrum of the source.
We downloaded the observation directory for observation ID 30002006002 from the HEASARC and used the FTOOL \texttt{barycorr} on the L2 cleaned science event files (file name ending with \verb|01_cl.evt|) to correct the photon arrival times to the solar system barycenter.
For our analysis, we considered photons from 3 to 79 keV at most 50\arcsec\ from the nominal position of the source, extracted from the two identical \textit{Focal Plane Modules A} and \textit{B} (FPMA and FPMB, respectively) onboard the spacecraft.
For this work, we used a total of $32.67\,\mathrm{ks}$ of good time intervals (GTIs), only selecting intervals longer than $10\,\mathrm{s}$. In Figure \ref{fig:herx1_lc} (upper panel), we present the light curve of the \nustar Her X-1 observations. The source varied substantially in brightness during the observations on fairly long time scales (10s to 1000s of seconds), indicating a significant source of red noise at low frequencies.

In the lower panel of Figure \ref{fig:herx1_lc}, we present an averaged periodogram created from a total of $3260$ light curve segments, each $10\,\mathrm{s}$ in duration. The periodogram shows very strong peaks at the $1.24\,\mathrm{s}$ rotational period of the pulsar, along with its first five harmonics. While at low frequencies, the deviation from the expected noise distribution (a $\chi^2$ distribution centered on $2$) can be explained with the longer-term variability within the source, at frequencies above $5 \,\mathrm{Hz}$ or so, the periodogram displays the typical oscillatory pattern associated with dead time (see also \citealt{Bachetti+15}), suggesting that the standard distributions usually applied to periodograms will not produce unbiased results for these observations.

\begin{figure*}
\begin{center}
\includegraphics[width=\textwidth]{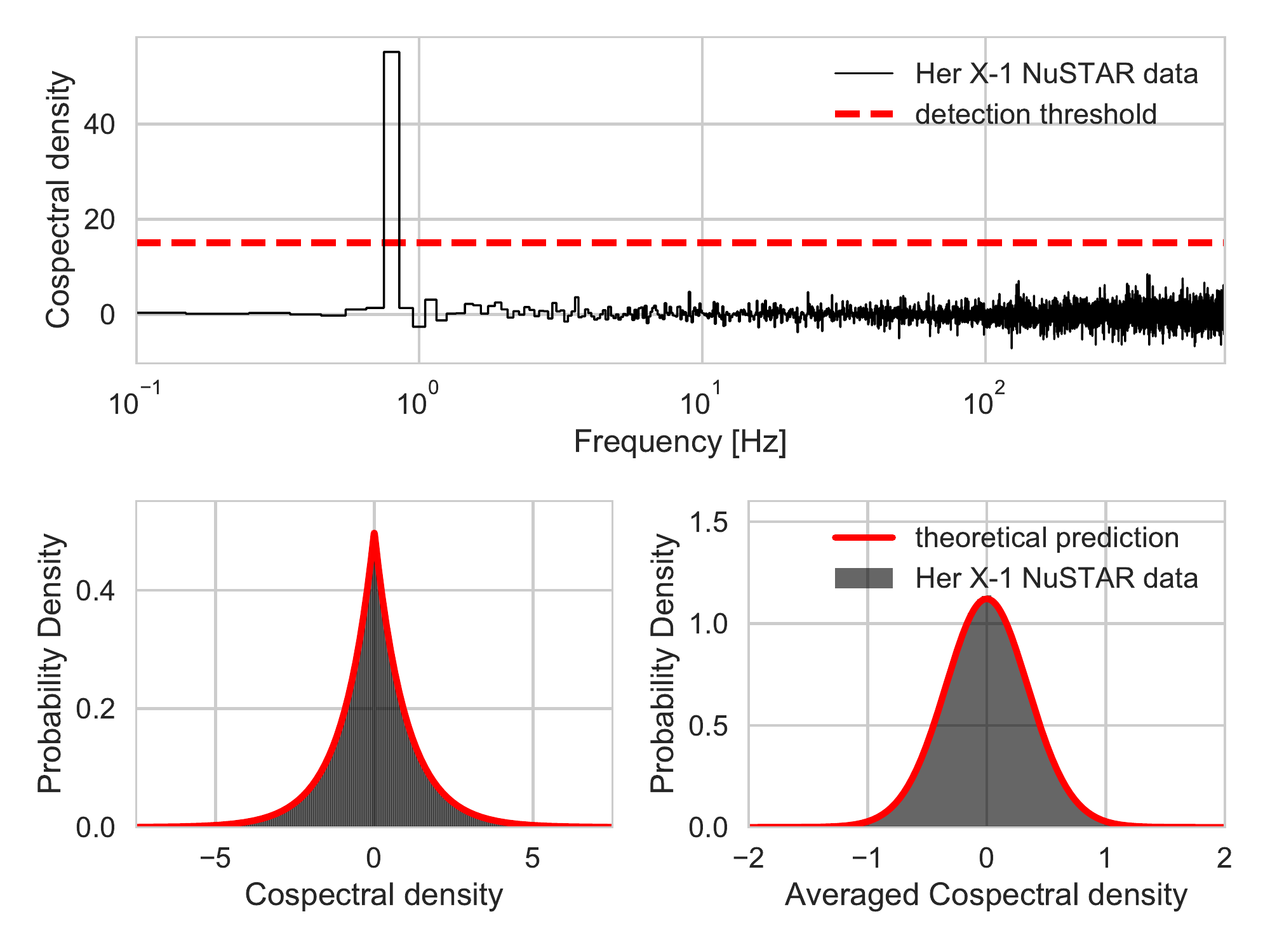}
\caption{Upper panel: Cospectrum of a $10\,\mathrm{s}$ segment of the \nustar\ observations of Her X-1 (black). The cospectrum is created by observing the same source simultaneously, but statistically independently with the two identical FPMA and FPMB detectors. The rotational signal of the pulsar appears very clearly at $\nu_\mathrm{rot} = 0.806\,\mathrm{Hz}$, but unlike the periodogram, the cospectrum is not affected by dead time. In red, we show the $p = 10^{-3}$ detection threshold, corrected for $6999$ trials (equivalent to the number of frequencies in the periodogram). Lower left:  histogram (grey) of the cospectral densities collected from 3260 individual cospectra between $50\,\mathrm{Hz}$ and $400\,\mathrm{Hz}$. In red, we also show the theoretically predicted Laplace distribution. Lower right: in black, the histogram of cospectral densities collected from $7$ averaged cospectra, each created from $15$ light curve segments of $200\,\mathrm{s}$ duration. Again, only cospectral densities between $50\,\mathrm{Hz}$ and $400\,\mathrm{Hz}$ were used in this figure. In red, we show the theoretical probability density function defined in Equation \ref{eqn:averaged_pdf} with $n = 15$. For both averaged and single cospectra, the observed histogram matches the theoretically expected distributions very closely.
}
\label{fig:herx1_cs}
\end{center}
\end{figure*}

In Figure \ref{fig:herx1_cs} (upper panel), we present a cospectrum of the Her X-1 \nustar\ observation. Because the periodic signal generated by the pulsar dominates the cospectrum, we plot the cospectrum of a short segment of $10\,\mathrm{s}$ duration starting at $\mathrm{MET} = 85740109.54169$ for illustrative purpose to highlight the noise properties. Unlike the periodogram, the cospectrum is not modulated by dead time, and the statistical distributions defined in Section \ref{sec:whitenoise_cospectra} apply once the cospectra are corrected using the FAD technique (see also Figure \ref{fig:herx1_cs}, lower left panel and below). We also show the trial-corrected detection threshold under the null hypothesis that the cospectrum consists solely of white noise. The highest cospectral density of $P = 55.16$ occurs at the frequency of the pulsar, $\nu_\mathrm{rot} = 0.806\,\mathrm{Hz}$; the probability of observing this power under the null hypothesis is effectively $p \sim 0$ within the limits of numerical accuracy. 

In order to test our theoretical predictions for the distributions governing (averaged) cospectra, we first produced $3260$ individual FAD-corrected cospectra out of each of the $10\,\mathrm{s}$ segments, and for each cospectrum extracted the cospectral densities in the range $50\,\mathrm{Hz}$ to $400\,\mathrm{Hz}$, where we do not expect strong contamination by the pulsar signal and its harmonics, or by the low-frequency red noise component.  We plot a histogram of cospectral densities between $50\,\mathrm{Hz}$ and $400\,\mathrm{Hz}$ for all $3260$ segments (a total of $11406500$ cospectral densities) in Figure \ref{fig:herx1_cs} (lower left panel), along with the theoretically expected Laplace distribution. We find a generally high agreement between the Laplace distribution and the distribution of observed cospectral densities. 

In order to repeat the same process for averaged cospectra, we take the $7$ GTIs that are longer than $3000\,\mathrm{s}$  and extract a light curve of exactly $3000\,\mathrm{s}$ from each. From each of these light curves, we produce an averaged cospectrum by averaging $15$ consecutive light curve segments of $200\,\mathrm{s}$ duration. This yields $7$ averaged cospectra. As above, we extract cospectral densities between $50\,\mathrm{Hz}$ and $400\,\mathrm{Hz}$ and plot the histogram of cospectral densities extracted from all averaged cospectra (910000 cospectral densitites) in Figure \ref{fig:herx1_cs} (lower right panel). As with the single cospectrum, the histogram for the averaged cospectral densities matches the theoretical prediction very closely. This indicates that there is no evidence in the data that would argue against the observed cospectral densities being drawn from the distribution in Equation \ref{eqn:averaged_cdf}.

\section{Discussion and Conclusions}
\label{sec:discussion}
We have derived the statistical distribution for the cospectrum, defined as the real part of the cross spectrum. We show that because the Fourier amplitudes being multiplied to derive the cospectrum are now no longer identical (as is the case in the periodogram), the statistical distributions no longer reduce to a simple $\chi^2$ distribution with two degrees of freedom. Instead, we find that the densities in a single cospectrum follow a Laplace distribution with a mean of $\mu=0$ and a width of $\sigma=1$. This has important consequences for period detection. Most importantly, the Laplace distribution is considerably narrower than the $\chi^2_2$ distribution expected for periodograms, and thus the significance of a candidate periodic signal will generally be underestimated when using the latter. Using the correct distribution therefore helps correctly identifying weak signals, which the $\chi^2_2$ distribution would ignore as false negatives. 
Similarly, we find that the sums of Laplace distributions do not follow a similarly simple expression as in the periodogram case, but is considerably more computationally expensive, and may be difficult to estimate numerically when the number $n$ of averaged light curves in the final cospectrum is large. However, we find that for $n \gtrsim 30$, the statistical distribution can be well approximated by a Gaussian distribution, and the resulting tail probabilities used for period detection are very nearly the same as those derived from the exact distribution, up to a tail probability of $p \approx 10^{-4}$. This conclusion, however, depends sensitively on the detection threshold as well as the number of trials: for very small tail probabilities, the two distributions may still deviate significantly. For practical purposes, we suggest using Equation \ref{eqn:averaged_cdf} for at least up to $\sim$30 averaged cospectra, but also for averages of more spectra if significance thresholds for tail probabilities are smaller than $10^{-4}$, or the number of trials is large. 

As an example, we have simulated how a QPO would appear in \nustar\ in the presence of dead time, and have shown that the shape of the periodogram is strongly distorted, whereas that of the cospectrum is not (for a longer introduction into the cospectrum and how it can be used in the presence of dead time, see also \citealt{Bachetti+15}). The significance of the QPO is difficult to assess in the periodogram, because of the non-linearities introduced by the variable dead time. The standard $\chi^2_2$ distributions may either overestimate or underestimate the significance, depending on the shape of the underlying power spectrum and its modification due to dead time at a given frequency, adding complexity to the detection process in the form of finding a non-linear model for the dead time. The cospectrum, on the other hand, only requires an estimate of the local variance in order to use the equations derived above, making it a far more convenient choice for periodicity detection. By applying a straightforward correction using the differences between Fourier amplitudes derived from two independent detectors, cospectra can be corrected such that they follow the distributions derived here very closely.

We show that the same is true for observations of the bright neutron star X-ray binary Her X-1, where the intrinsic brightness of the source leads to strong modulations in the periodogram due to dead time. For sources observed with \nustar\ and other instruments with a similar set-up of at least two redundant, identical detectors, one can take advantage of the independent, simultaneous observations in two detectors to form a cospectrum. While any (quasi-)periodic signal will appear weaker in the cospectrum than in the periodogram (even in the case, as presented here, where the signals in the two detectors are exactly in phase), it has the substantial advantage that it is less affected by dead time and other similar detector effects. We show that the cospectrum of the observations observed with \nustar\ follow the expected theoretical distributions very closely once the FAD correction is applied.

The distributions laid out above allow for detecting periodic and narrow quasi-periodic signals in the presence of detector white noise, and especially important in the context of pulsar detection in X-rays, where faint sources may yield marginal detections even in the best of cases. At the same time, as instruments like \astrosat\ and \ixpe\ allow for observations with higher sensitivity, incorporating an accurate treatment of instrumental biases becomes increasingly important, and the cospectral statistics laid out here provide powerful tools to do so.

Notably absent from this discussion, however, is the much more common case where a source exhibits stochastic variability in the form of red noise or notably broadened quasi-periodic oscillations. In this case, the goal is either estimation of the precise properties of the underlying stochastic process, or detection of periods against a background varying stochastically. As has been shown above, the fact that the cross spectrum consists of two different time series complicates the statistical distributions considerably, and this is similarly true cospectra with variability. The exact treatment of this case is beyond the scope of this paper, and will be considered in depth in a forthcoming publication.

\paragraph{Acknowledgements}
The authors thank the anonymous referee for their helpful comments.
The authors also thank Thomas Laetsch for helpful suggestions regarding the mathematical derivations, and Peter Bult for spotting a mathematical error.
DH is supported by the James Arthur Postdoctoral Fellowship and the Moore-Sloan Data Science Environment at New York University. 
DH acknowledges support from the DIRAC Institute in the Department of Astronomy at the University of Washington. The DIRAC Institute is supported through generous gifts from the Charles and Lisa Simonyi Fund for Arts and Sciences, and the Washington Research Foundation.
MB is supported in part by the Italian Space Agency through agreement ASI-INAF n.2017-12-H.0 and ASI-INFN agreement n.2017-13-H.0.



\bibliography{cospectra-paper}

\begin{thebibliography}{}
\expandafter\ifx\csname natexlab\endcsname\relax\def\natexlab#1{#1}\fi

\bibitem[{{Anderson} {et~al.}(1994){Anderson}, {Wachter}, {Margon}, {Downes},
  {Blair}, \& {Halpern}}]{anderson1994}
{Anderson}, S.~F., {Wachter}, S., {Margon}, B., {et~al.} 1994, \apj, 436, 319

\bibitem[{{Ardhuin} {et~al.}(2016){Ardhuin}, {Sutherland}, {Doble}, \&
  {Wadhams}}]{ardhuin2016}
{Ardhuin}, F., {Sutherland}, P., {Doble}, M., \& {Wadhams}, P. 2016, \grl, 43,
  5775

\bibitem[{{Bachetti}(2015)}]{bachetti2015b}
{Bachetti}, M. 2015, {MaLTPyNT: Quick look timing analysis for NuSTAR data},
  Astrophysics Source Code Library, ascl:1502.021

\bibitem[{{Bachetti}(2016)}]{bachetti2016}
---. 2016, Astronomische Nachrichten, 337, 349

\bibitem[{{Bachetti} \& {Huppenkothen}(2017)}]{bachetti2017}
{Bachetti}, M., \& {Huppenkothen}, D. 2017, ArXiv e-prints, arXiv:1709.09700

\bibitem[{Bachetti {et~al.}(2015)Bachetti, Harrison, Cook, Tomsick, Schmid,
  Grefenstette, Barret, Boggs, Christensen, Craig, Fabian, F{\"u}rst, Gandhi,
  Hailey, Kara, Maccarone, Miller, Pottschmidt, Stern, Uttley, Walton, Wilms,
  \& Zhang}]{Bachetti+15}
Bachetti, M., Harrison, F.~A., Cook, R., {et~al.} 2015, ApJ, 800, 109

\bibitem[{{Bahcall} \& {Bahcall}(1972)}]{bahcall1972}
{Bahcall}, J.~N., \& {Bahcall}, N.~A. 1972, \apjl, 178, L1

\bibitem[{Balakrishnan \& Kocherlakota(1986)}]{balakrishnan1986}
Balakrishnan, N., \& Kocherlakota, S. 1986, Sankhy?: The Indian Journal of
  Statistics, Series B (1960-2002), 48, 439

\bibitem[{{Barri{\`e}re} {et~al.}(2015){Barri{\`e}re}, {Krivonos}, {Tomsick},
  {Bachetti}, {Boggs}, {Chakrabarty}, {Christensen}, {Craig}, {Hailey},
  {Harrison}, {Hong}, {Mori}, {Stern}, \& {Zhang}}]{barriere2015}
{Barri{\`e}re}, N.~M., {Krivonos}, R., {Tomsick}, J.~A., {et~al.} 2015, \apj,
  799, 123

\bibitem[{{Cheng} {et~al.}(1995){Cheng}, {Vrtilek}, \& {Raymond}}]{cheng1995}
{Cheng}, F.~H., {Vrtilek}, S.~D., \& {Raymond}, J.~C. 1995, \apj, 452, 825

\bibitem[{{Davidsen} {et~al.}(1972){Davidsen}, {Henry}, {Middleditch}, \&
  {Smith}}]{davidsen1972}
{Davidsen}, A., {Henry}, J.~P., {Middleditch}, J., \& {Smith}, H.~E. 1972,
  \apjl, 177, L97

\bibitem[{{Epitropakis} \& {Papadakis}(2016)}]{epitropakis2016}
{Epitropakis}, A., \& {Papadakis}, I.~E. 2016, \aap, 591, A113

\bibitem[{{Ferrigno} {et~al.}(2017){Ferrigno}, {Bozzo}, {Sanna}, {Pintore},
  {Papitto}, {Riggio}, {Burderi}, {Di Salvo}, {Iaria}, \&
  {D'A{\`i}}}]{ferrigno2017}
{Ferrigno}, C., {Bozzo}, E., {Sanna}, A., {et~al.} 2017, \mnras, 466, 3450

\bibitem[{{Fiedler} {et~al.}(2015){Fiedler}, {Brodie}, {McNinch}, \&
  {Guza}}]{fiedler2015}
{Fiedler}, J.~W., {Brodie}, K.~L., {McNinch}, J.~E., \& {Guza}, R.~T. 2015,
  \grl, 42, 9933

\bibitem[{{Forman} {et~al.}(1972){Forman}, {Jones}, \& {Liller}}]{forman1972}
{Forman}, W., {Jones}, C.~A., \& {Liller}, W. 1972, \apjl, 177, L103

\bibitem[{F{\"u}rst {et~al.}(2013)F{\"u}rst, Grefenstette, Staubert, Tomsick,
  Bachetti, Barret, Bellm, Boggs, Chenevez, Christensen, Craig, Hailey,
  Harrison, Klochkov, Madsen, Pottschmidt, Stern, Walton, Wilms, \&
  Zhang}]{fuerst13}
F{\"u}rst, F., Grefenstette, B.~W., Staubert, R., {et~al.} 2013, The
  Astrophysical Journal, 779, 69

\bibitem[{{Giacconi} {et~al.}(1973){Giacconi}, {Gursky}, {Kellogg}, {Levinson},
  {Schreier}, \& {Tananbaum}}]{giacconi1973}
{Giacconi}, R., {Gursky}, H., {Kellogg}, E., {et~al.} 1973, \apj, 184, 227

\bibitem[{{Gregory} \& {Loredo}(1992)}]{gregory1992}
{Gregory}, P.~C., \& {Loredo}, T.~J. 1992, \apj, 398, 146

\bibitem[{{Groth}(1975)}]{groth1975}
{Groth}, E.~J. 1975, \apjs, 29, 285

\bibitem[{{Huppenkothen} {et~al.}(2013){Huppenkothen}, {Watts}, {Uttley}, {van
  der Horst}, {van der Klis}, {Kouveliotou}, {G{\"o}{\v g}{\"u}{\c s}},
  {Granot}, {Vaughan}, \& {Finger}}]{huppenkothen2013}
{Huppenkothen}, D., {Watts}, A.~L., {Uttley}, P., {et~al.} 2013, \apj, 768, 87

\bibitem[{{Huppenkothen} {et~al.}(2017){Huppenkothen}, {Younes}, {Ingram},
  {Kouveliotou}, {G{\"o}{\u g}{\"u}{\c s}}, {Bachetti},
  {S{\'a}nchez-Fern{\'a}ndez}, {Chenevez}, {Motta}, {van der Klis}, {Granot},
  {Gehrels}, {Kuulkers}, {Tomsick}, \& {Walton}}]{huppenkothen2017}
{Huppenkothen}, D., {Younes}, G., {Ingram}, A., {et~al.} 2017, \apj, 834, 90

\bibitem[{{Igna} \& {Leahy}(2011)}]{igna2011}
{Igna}, C.~D., \& {Leahy}, D.~A. 2011, \mnras, 418, 2283

\bibitem[{{Ingram} {et~al.}(2016){Ingram}, {van der Klis}, {Middleton}, {Done},
  {Altamirano}, {Heil}, {Uttley}, \& {Axelsson}}]{ingram2016}
{Ingram}, A., {van der Klis}, M., {Middleton}, M., {et~al.} 2016, \mnras, 461,
  1967

\bibitem[{{John} \& {Kishore Kumar}(2016)}]{john2016}
{John}, S.~R., \& {Kishore Kumar}, K. 2016, Journal of Atmospheric and
  Solar-Terrestrial Physics, 138, 74

\bibitem[{Kotz {et~al.}(2001)Kotz, Kozubowski, \& Podg{\'o}rski}]{kotz2001}
Kotz, S., Kozubowski, T.~J., \& Podg{\'o}rski, K. 2001, Asymmetric Multivariate
  Laplace Distribution (Boston, MA: Birkh{\"a}user Boston), 239--272

\bibitem[{{Leahy} \& {Abdallah}(2014)}]{leahy2014}
{Leahy}, D.~A., \& {Abdallah}, M.~H. 2014, \apj, 793, 79

\bibitem[{{Leahy} {et~al.}(1983){Leahy}, {Darbro}, {Elsner}, {Weisskopf},
  {Kahn}, {Sutherland}, \& {Grindlay}}]{leahy1983}
{Leahy}, D.~A., {Darbro}, W., {Elsner}, R.~F., {et~al.} 1983, \apj, 266, 160

\bibitem[{{Lomb}(1976)}]{lomb1976}
{Lomb}, N.~R. 1976, \apss, 39, 447

\bibitem[{{Nowak} {et~al.}(1999){Nowak}, {Vaughan}, {Wilms}, {Dove}, \&
  {Begelman}}]{nowak1999}
{Nowak}, M.~A., {Vaughan}, B.~A., {Wilms}, J., {Dove}, J.~B., \& {Begelman},
  M.~C. 1999, \apj, 510, 874

\bibitem[{{Paterna} {et~al.}(2016){Paterna}, {Crivelli}, \&
  {Lehning}}]{paterna2016}
{Paterna}, E., {Crivelli}, P., \& {Lehning}, M. 2016, \grl, 43, 4441

\bibitem[{{Reynolds} {et~al.}(1997){Reynolds}, {Quaintrell}, {Still}, {Roche},
  {Chakrabarty}, \& {Levine}}]{reynolds1997}
{Reynolds}, A.~P., {Quaintrell}, H., {Still}, M.~D., {et~al.} 1997, \mnras,
  288, 43

\bibitem[{{Scargle}(1982)}]{scargle1982}
{Scargle}, J.~D. 1982, \apj, 263, 835

\bibitem[{{Scott} \& {Leahy}(1999)}]{scott1999}
{Scott}, D.~M., \& {Leahy}, D.~A. 1999, \apj, 510, 974

\bibitem[{Seijas-Mac{\'\i}as \& Oliveira(2012)}]{seijas2012}
Seijas-Mac{\'\i}as, A., \& Oliveira, A. 2012, Discussiones Mathematicae
  Probability and Statistics, 32, 87

\bibitem[{{Stiele} \& {Kong}(2017)}]{stiele2017}
{Stiele}, H., \& {Kong}, A.~K.~H. 2017, \apj, 844, 8

\bibitem[{{Tananbaum} {et~al.}(1972){Tananbaum}, {Gursky}, {Kellogg},
  {Levinson}, {Schreier}, \& {Giacconi}}]{tananbaum1972}
{Tananbaum}, H., {Gursky}, H., {Kellogg}, E.~M., {et~al.} 1972, \apjl, 174,
  L143

\bibitem[{{Uttley} {et~al.}(2014){Uttley}, {Cackett}, {Fabian}, {Kara}, \&
  {Wilkins}}]{uttley2014}
{Uttley}, P., {Cackett}, E.~M., {Fabian}, A.~C., {Kara}, E., \& {Wilkins},
  D.~R. 2014, \aapr, 22, 72

\bibitem[{{van der Klis}(1989)}]{vanderklis1989}
{van der Klis}, M. 1989, in NATO Advanced Science Institutes (ASI) Series C,
  Vol. 262, NATO Advanced Science Institutes (ASI) Series C, ed.
  H.~{{\"O}gelman} \& E.~P.~J. {van den Heuvel}, 27

\bibitem[{{Vaughan} \& {Nowak}(1997)}]{vaughan1997}
{Vaughan}, B.~A., \& {Nowak}, M.~A. 1997, \apjl, 474, L43

\bibitem[{{Vigeesh} {et~al.}(2017){Vigeesh}, {Jackiewicz}, \&
  {Steiner}}]{vigeesh2017}
{Vigeesh}, G., {Jackiewicz}, J., \& {Steiner}, O. 2017, \apj, 835, 148

\bibitem[{{Wang} \& {Toigo}(2016)}]{wang2016}
{Wang}, H., \& {Toigo}, A.~D. 2016, \icarus, 271, 207

\bibitem[{{Wang} \& {Nakamura}(2015)}]{wang2015}
{Wang}, L., \& {Nakamura}, N. 2015, \grl, 42, 8192

\bibitem[{Watson(1922)}]{watson1922}
Watson, G.~N. 1922, A treatise on the theory of Bessel functions (Cambridge,
  University Press)

\bibitem[{Wishart \& Bartlett(1932)}]{wishart1932}
Wishart, J., \& Bartlett, M.~S. 1932, Mathematical Proceedings of the Cambridge
  Philosophical Society, 28, 455?459

\bibitem[{Zhang {et~al.}(1995)Zhang, Jahoda, Swank, Morgan, \&
  Giles}]{Zhang+95}
Zhang, W., Jahoda, K., Swank, J.~H., Morgan, E.~H., \& Giles, A.~B. 1995, ApJ,
  449, 930

\bibitem[{{Zoghbi} {et~al.}(2016){Zoghbi}, {Miller}, {King}, {Miller}, {Proga},
  {Kallman}, {Fabian}, {Harrison}, {Kaastra}, {Raymond}, {Reynolds}, {Boggs},
  {Christensen}, {Craig}, {Hailey}, {Stern}, \& {Zhang}}]{zoghbi2016}
{Zoghbi}, A., {Miller}, J.~M., {King}, A.~L., {et~al.} 2016, \apj, 833, 165

\bibitem[{{Zurita-Gotor}(2017)}]{zurita-gotor2017}
{Zurita-Gotor}, P. 2017, \grl, 44, 2007

\end{thebibliography}
\bibliographystyle{apj}


\end{document}